\documentclass{article}

\usepackage[nonatbib,final]{nips_2016}

\usepackage[utf8]{inputenc} % allow utf-8 input
\usepackage[T1]{fontenc}    % use 8-bit T1 fonts
\usepackage{hyperref}       % hyperlinks
\usepackage{url}            % simple URL typesetting
\usepackage{booktabs}       % professional-quality tables
\usepackage{amsfonts}       % blackboard math symbols
\usepackage{nicefrac}       % compact symbols for 1/2, etc.
\usepackage{microtype}      % microtypography

\usepackage{float}
\usepackage{graphicx}
\usepackage{amsmath}
\usepackage{caption}
\usepackage{subcaption}
\usepackage{color}

\newcommand{\be}{\begin{equation}}
\newcommand{\ee}{\end{equation}}
\newcommand{\ba}{\begin{align}}
\newcommand{\ea}{\end{align}}
\newcommand{\bea}{\begin{eqnarray}}
\newcommand{\eea}{\end{eqnarray}}
\newcommand{\lpr}{\left(}
\newcommand{\rpr}{\right)}

\newcommand{\vx}{{\boldsymbol{x}}}

\newcommand{\vtheta}{{\boldsymbol{\theta}}}

\newcommand{\vphi}{{\boldsymbol{\phi}}}

\newcommand{\vTheta}{{\boldsymbol{\Theta}}}

\newcommand{\br}{{\textbf{r}}}

\newcommand{\bb}{{\textbf{b}}}

\newcommand{\bA}{{\textbf{A}}}

\newcommand{\bg}{{\textbf{g}}}

\newcommand{\bG}{{\textbf{G}}}
\newcommand{\bU}{{\textbf{U}}}

\title{Inference by Reparameterization in Neural Population Codes}

\author{
  Rajkumar V.~Raju\\
  Department of ECE\\
  Rice University\\
  Houston, Tx 77005 \\
  \texttt{rv12@rice.edu}\\
   \And
   Xaq Pitkow \\
   Dept. of Neuroscience, Dept. of ECE\\
   Baylor College of Medicine,   Rice University\\
   Houston, Tx 77005 \\
   \texttt{xaq@rice.edu} \\
}

\begin{document}

\maketitle

\begin{abstract}

Behavioral experiments on humans and animals suggest that the brain performs probabilistic inference to interpret its environment. Here we present a new general-purpose, biologically-plausible neural implementation of approximate inference. The neural network represents uncertainty using Probabilistic Population Codes (PPCs), which are distributed neural representations that naturally encode probability distributions, and support marginalization and evidence integration in a biologically-plausible manner. By connecting multiple PPCs together as a probabilistic graphical model, we represent multivariate probability distributions. Approximate inference in graphical models can be accomplished by message-passing algorithms that disseminate local information throughout the graph. An attractive and often accurate example of such an algorithm is Loopy Belief Propagation (LBP), which uses local marginalization and evidence integration operations to perform approximate inference efficiently even for complex models. Unfortunately, a subtle feature of LBP renders it neurally implausible. However, LBP can be elegantly reformulated as a sequence of Tree-based Reparameterizations (TRP) of the graphical model. We re-express the TRP updates as a nonlinear dynamical system with both fast and slow timescales, and show that this produces a neurally plausible solution. By combining all of these ideas, we show that a network of PPCs can represent multivariate probability distributions and implement the TRP updates to perform probabilistic inference. Simulations with Gaussian graphical models demonstrate that the neural network inference quality is comparable to the direct evaluation of LBP and robust to noise, and thus provides a promising mechanism for general probabilistic inference in the population codes of the brain.

\end{abstract}

\section{Introduction}

In everyday life we constantly face tasks we must perform in the presence of sensory uncertainty. A natural and efficient strategy is then to use probabilistic computation. Behavioral experiments have established that humans and animals do in fact use probabilistic rules in sensory, motor and cognitive domains \cite{knill1996perception,doya2007bayesian,pouget2013probabilistic}. However, the implementation of such computations at the level of neural circuits is not well understood.

In this work, we ask how distributed neural computations can consolidate incoming sensory information and reformat it so it is accessible for many tasks. More precisely, how can the brain simultaneously infer marginal probabilities in a probabilistic model of the world? Previous efforts to model marginalization in neural networks using distributed codes invoked limiting assumptions, either treating only a small number of variables \cite{beck2011marginalization}, allowing only binary variables \cite{ott2006neurodynamics,steimer2009belief,litvak2009cortical}, or restricting interactions \cite{george2009towards,grabska2013demixing}. Real-life tasks are more complicated and involve a large number of variables that need to be marginalized out, requiring a more general inference architecture.

Here we present a distributed, nonlinear, recurrent network of neurons that performs inference about many interacting variables. There are two crucial parts to this model: the representation and the inference algorithm. We assume that brains represent probabilities over individual variables using Probabilistic Population Codes (PPCs) \cite{ma2006bayesian}, which were derived from using Bayes' Rule on experimentally measured neural responses to sensory stimuli. Here for the first time we link multiple PPCs together to construct a large-scale graphical model. For the inference algorithm, many researchers have considered Loopy Belief Propagation (LBP) to be a simple and efficient candidate algorithm for the brain \cite{pearl1988probabilistic,yedidia2003understanding,lee2003hierarchical,rao2004hierarchical,george2009towards,ott2006neurodynamics,litvak2009cortical,steimer2009belief}. However, we will discuss one particular feature of LBP that makes it neurally implausible. Instead, we propose that an alternative formulation of LBP known as Tree-based Reparameterization (TRP) \cite{wainwright2003tree}, with some modifications for continuous-time operation at two timescales, is well-suited for neural implementation in population codes.

We describe this network mathematically below, but the main conceptual ideas are fairly straightforward: multiplexed patterns of activity encode statistical information about subsets of variables, and neural interactions disseminate these statistics to all other encoded variables for which they are relevant.

In Section \ref{PPC} we review key properties of our model of how neurons can represent probabilistic information through Probabilistic Population Codes. Section \ref{TRP} reviews graphical models, Loopy Belief Propagation, and Tree-based Reparameterization. In Section \ref{sec:NeuralTRP}, we merge these ingredients to model how populations of neurons can represent and perform inference on large multivariate distributions. Section \ref{Experiments} describes experiments to test the performance of network. We summarize and discuss our results in Section \ref{Conclusion}.

\section{Probabilistic Population Codes}
\label{PPC}

Neural responses $\br$ vary from trial to trial, even to repeated presentations of the same stimulus $x$. This variability can be expressed as the likelihood function $p(\br|x)$. Experimental data from several brain areas responding to simple stimuli suggests that this variability often belongs to the exponential family of distributions with linear sufficient statistics \cite{ma2006bayesian, jazayeri2006optimal, beck2008probabilistic, beck2011marginalization, graf2011decoding}:
\begin{equation}
p(\br|x) = \phi(\br)\exp(\textbf{h}(x) \cdot \br),
\end{equation}
where $\textbf{h}(x)$ depends on the stimulus-dependent mean and fluctuations of the neuronal response and $\phi(\br)$ is independent of the stimulus. For a conjugate prior $p(x)$, the posterior distribution will also have this general form, $p(x|\br) \propto \exp(\textbf{h}(x) \cdot \br)$. This neural code is known as a linear PPC: it is a Probabilistic Population Code because the population activity collectively encodes the stimulus probability, and it is linear because the log-likelihood is linear in $\br$. In this paper, we assume responses are drawn from this family, although incorporation of more general PPCs with nonlinear sufficient statistics $\textbf{T}(\br)$ is possible: $p(\br|x) \propto \exp(\textbf{h}(x) \cdot \textbf{T}(\br))$.

%\subsection{Key Properties of PPCs}

\begin{figure}[h]
	\centering
	\includegraphics[width=0.7\textwidth]{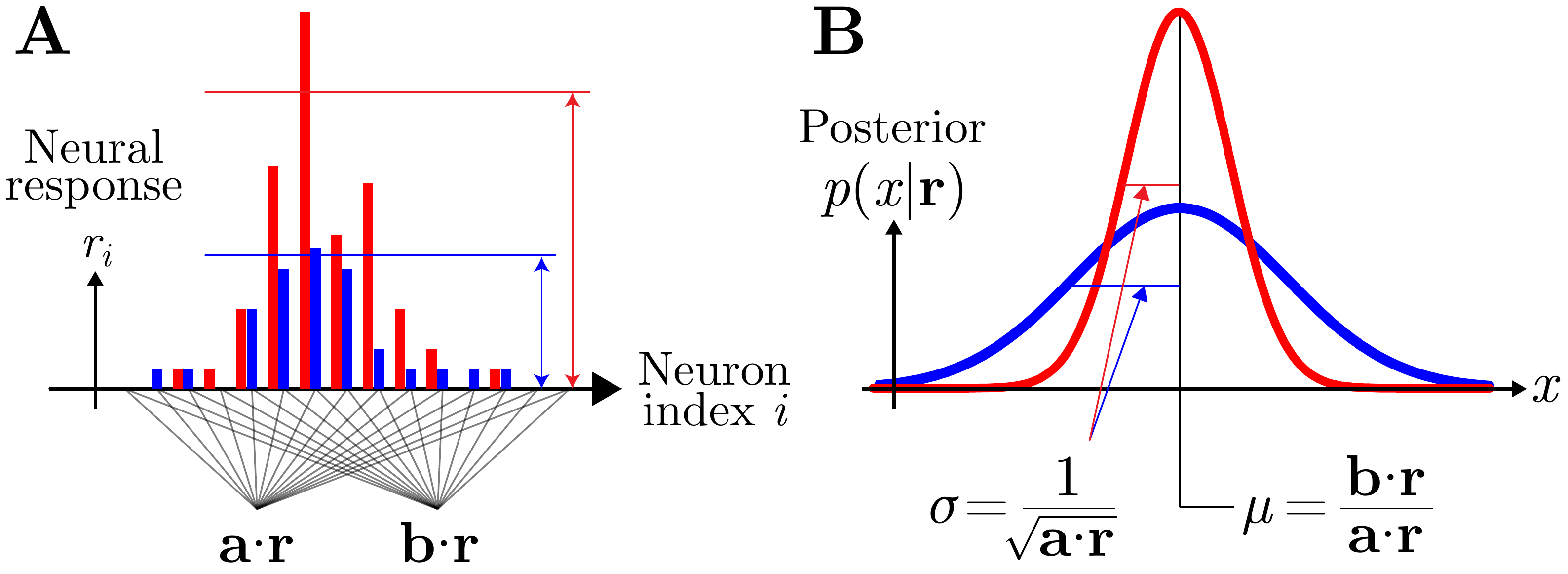}
	\caption{Key properties of linear PPCs. ({\bf A}) Two single trial population responses for a particular stimulus, with low and high amplitudes (blue and red). The two projections $\textbf{a}\cdot\br$ and $\textbf{b}\cdot\br$ encode the natural parameters of the posterior. ({\bf B}) Corresponding posteriors over stimulus variables determined by the responses in panel A. The gain or overall amplitude of the population code is inversely proportional to the variance of the posterior distribution.}
	\label{fig:PPCFig}
\end{figure}

An important property of linear PPCs, central to this work, is that different projections of the population activity encode the natural parameters of the underlying posterior distribution. For example, if the posterior distribution is Gaussian (Figure \ref{fig:PPCFig}), then
$p(x|\br) \propto \exp{\left(-\tfrac{1}{2}x^2\textbf{a}\cdot\br + x \textbf{b}\cdot\br\right)}$,
with $\textbf{a}\cdot\br$ and $\textbf{b}\cdot\br$ encoding the linear and quadratic natural parameters of the posterior. These projections are related to the expectation parameters, the mean and variance, by $\mu = \frac{\textbf{b}\cdot\br}{\textbf{a}\cdot\br}$ and $\sigma^2 = \frac{1}{\textbf{a}\cdot\br}$.

A second important property of linear PPCs is that the variance of the encoded distribution is inversely proportional to the overall amplitude of the neural activity. Intuitively, this means that more spikes means more certainty (Figure \ref{fig:PPCFig}).

%\subsection{Basic Probabilistic Operations with PPCs}

The most fundamental probabilistic operations are the product rule and the sum rule. Linear PPCs can perform both of these operations while maintaining a consistent representation \cite{beck2011marginalization}, a useful feature for constructing a model of canonical computation. For a log-linear probability code like linear PPCs, the product rule corresponds to weighted summation of neural activities: $p(x|\br_1,\br_2)\propto p(x|\br_1)p(x|\br_2) \Longleftrightarrow \br_3=A_1\br_1+A_2\br_2$. In contrast, to use the sum rule to marginalize out variables, linear PPCs require nonlinear transformations of population activity. Specifically, a quadratic nonlinearity with divisive normalization performs near-optimal marginalization in linear PPCs \cite{beck2011marginalization}. Quadratic interactions arise naturally through coincidence detection, and divisive normalization is a nonlinear inhibitory effect widely observed in neural circuits \cite{heeger1992normalization,carandini2012normalization,rubin2015stabilized}. Alternatively, near-optimal marginalizations on PPCs can also be performed by more general nonlinear transformations \cite{vasudeva2015marginalization}. In sum, PPCs provide a biologically compatible representation of probabilistic information.

\section{Inference by Tree-based Reparameterization}
\label{TRP}
%In this section, we introduce TRP updates for inference in a probabilistic graphical model.

\subsection{Graphical Models}
\label{GMs}

To generalize PPCs, we need to represent the joint probability distribution of many variables. A natural way to represent multivariate distributions is with probabilistic graphical models. In this work, we use the formalism of factor graphs, a type of bipartite graph in which nodes representing variables are connected to other nodes called factors representing interactions between `cliques' or sets of variables (Figure \ref{fig:TRP2}A). The joint probability over all variables can then be represented as a product over cliques, $p(\vx) = \frac{1}{Z}\prod_{c\in C} \psi_c(\vx_c)$, where $\psi_c(\vx_c)$ are nonnegative compatibility functions on the set of variables $\vx_c=\{x_c|c\in C\}$ in the clique, and $Z$ is a normalization constant. The distribution of interest will be a posterior distribution $p(\vx|\br)$ that depends on neural responses $\br$. Since the inference algorithm we present is unchanged with this conditioning, for notational convenience we suppress this dependence on $\br$.

In this paper, we focus on pairwise interactions, although our main framework generalizes naturally to richer, higher-order interactions. In a pairwise model, we allow singleton factors $\psi_s$ for variable nodes $s$ in a set of vertices $V$, and pairwise interaction factors $\psi_{st}$ for pairs $(s,t)$ in the set of edges $E$ that connect those vertices. The joint distribution is then
\begin{equation}
p(\vx) = \frac{1}{Z}\prod_{s\in V} \psi_s(x_s)\!\!\prod_{(s,t)\in E}\!\! \psi_{st}(x_s,x_t)
\label{pMRF}
\end{equation}

\subsection{Belief Propagation and its neural plausibility}
\label{BP}

The inference problem of interest in this work is to compute the marginal distribution for each variable, $p_s(x_s) = \int_{\vx\backslash x_s}p(\vx)\,d(\vx\backslash x_s)$. This task is generally intractable. However, the factorization structure of the distribution can be used to perform inference efficiently, either exactly in the case of tree graphs, or approximately for graphs with cycles. One such algorithm is called Belief Propagation (BP) \cite{pearl1988probabilistic}. BP iteratively passes information along the graph in the form of messages $m_{st}(x_t)$ from node $s$ to $t$, using only local computations that summarize the relevant aspects of other messages upstream in the graph:
\begin{equation}
m_{st}^{n+1} (x_t) = \int_{x_s}\!\! dx_s\,\psi_s(x_s) \psi_{st}(x_s,x_t)\!\!\!\!  \prod_{u \in N(s) \backslash t}\!\!\!\! m_{us}^n(x_s)
\hspace{3em}
b_s(x_s) \propto \psi_s\!\!\!\! \prod_{u \in N(s)} \!\!\!\! m_{us}(x_s)
\label{belief}
\end{equation}
where $n$ is the time or iteration number, and $N(s)$ is the set of neighbors of node $s$ on the graph. The estimated marginal, called the `belief' $b_s(x_s)$ at a node $s$, is proportional to the local evidence at that node $\psi_s(x_s)$ and all the messages coming into node $s$. Similarly, the messages themselves are determined self-consistently by combining incoming messages --- except for the previous message from the target node $t$.

This message exclusion is critical because it prevents evidence previously passed by the target node from being counted as if it were new evidence. This exclusion only prevents overcounting on a tree graph, and is unable to prevent overcounting of evidence passed around loops. For this reason, BP is exact for trees, but only approximate for general, loopy graphs. If we use this algorithm anyway, it is called `Loopy' Belief Propagation (LBP), and it often has quite good performance \cite{yedidia2003understanding}.

%\subsection{Neural plausibility of BP}

Multiple researchers have been intrigued by the possibility that the brain may perform LBP \cite{lee2003hierarchical,rao2004hierarchical,ott2006neurodynamics,george2009towards,litvak2009cortical,steimer2009belief}, since it gives ``a principled framework for propagating, in parallel, information and uncertainty between nodes in a network'' \cite{yedidia2003understanding}. Despite the conceptual appeal of LBP, it is important to get certain details correct: in an inference algorithm described by nonlinear dynamics, deviations from ideal behavior could in principle lead to very different outcomes.

One critically important detail is that each node must send different messages to different targets to prevent overcounting. This exclusion can render LBP neurally implausible, because neurons cannot readily send different output signals to many different target neurons. Some past work simply ignores the problem \cite{ott2006neurodynamics,litvak2009cortical}; the resultant overcounting destroys much of the inferential power of LBP, often performing worse than more na{\"i}ve algorithms like mean-field inference. %\cite{pitkow2010TAP}
One better option is to use different readouts of population activity for different targets \cite{steimer2009belief}, but this approach is inefficient because it requires many readout populations for messages that differ only slightly, and requires separate optimization for each possible target. Other efforts have avoided the problem entirely by performing only unidirectional inference of low-dimensional variables that evolve over time \cite{rao2004hierarchical}. Appealingly, one can circumvent all of these difficulties by using an alternative formulation of LBP known as Tree-based Reparameterization (TRP).

\subsection{Tree-based Reparameterization}

\begin{figure}[h]
\centering
\includegraphics[width=1\textwidth]{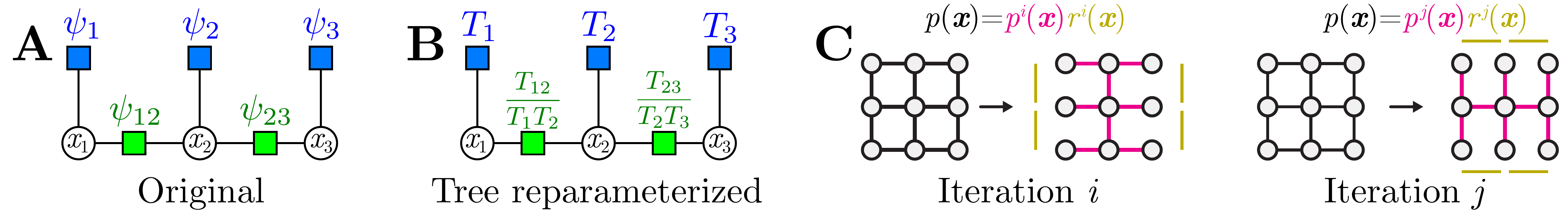}
\caption{Visualization of tree reparameterization. ({\bf A}) A probability distribution is specified by factors $\{\psi_s, \psi_{st}\}$ on a tree graph. ({\bf B}) An alternative parameterization of the same distribution in terms of the marginals $\{T_s, T_{st}\}$. ({\bf C}) Two TRP updates for a $3\times 3$ nearest-neighbor grid of variables.}
\label{fig:TRP2}
\end{figure}

Insightful work by Wainwright, Jakkola, and Willsky \cite{wainwright2003tree} revealed that belief propagation can be understood as a convenient way of refactorizing a joint probability distribution, according to approximations of local marginal probabilities. For pairwise interactions, this can be written as
\begin{equation}
p(\vx) \ \ = \ \ \frac{1}{Z}\prod_{s\in V} \psi_s(x_s)\!\!\!\prod_{(s,t)\in E} \!\!\!\psi_{st}(x_s,x_t)
\ \ =\ \ \prod_{s\in V}T_s(x_s)\!\!\! \prod_{(s,t) \in E}\!\! \frac{T_{st}(x_s,x_t)}{T_s(x_s)T_t(x_t)}
\label{TRPEqn}
\end{equation}
where $T_s(x_s)$ is a so-called `pseudomarginal' distribution of $x_s$ and $T_{st}(x_s,x_t)$ is a joint pseudomarginal over $x_s$ and $x_t$ (Figure \ref{fig:TRP2}A--B), where $T_s$ and $T_{st}$ are the outcome of Loopy Belief Propagation. The name pseudomarginal comes from the fact that these quantities are always locally consistent with being marginal distributions, but they are only globally consistent with the true marginals when the graphical model is tree-structured.

These pseudomarginals can be constructed iteratively as the true marginals of a different joint distribution $p^\tau(\vx)$ on an isolated tree-structured subgraph $\tau$. Compatibility functions $\psi$ from factors remaining outside of the subgraph are collected in a residual term $r^\tau(\vx)$. This regrouping leaves the joint distribution unchanged:
\begin{equation}
p(\vx) = p^\tau(\vx)r^\tau(\vx)
\end{equation}
The factors of $p^\tau$ are then rearranged by computing the true marginals on its subgraph $\tau$, again preserving the joint distribution. In subsequent updates, we iteratively refactorize using the marginals of $p^\tau$ along different tree subgraphs $\tau$ (Figure \ref{fig:TRP2}C).

Typical LBP can be interpreted as a sequence of local reparameterizations over just two neighboring nodes and their corresponding edge \cite{wainwright2003tree}. Pseudomarginals are initialized at time $n=0$ using the original factors: $T_s^0(x_s) \propto \psi_s(x_s)$ and $T_{st}^0(x_s,x_t) \propto \psi_s(x_s)\psi_t(x_t)\psi_{st}(x_s,x_t)$. At iteration $n+1$, the node and edge pseudomarginals are computed by exactly marginalizing the distribution built from previous pseudomarginals at iteration $n$:
\begin{equation}
\label{BPTRP}
T_s^{n+1} \propto T_s^{n}\!\!\! \prod_{u\in N(s)}\!\frac{1}{T_s^{n}} \int \! T_{su}^{n}\,dx_u
\hspace{3em}
T_{st}^{n+1} \propto \frac{T_{st}^{n}}{\left(\int \!{T_{st}^{n}\,dx_t}\right) \left(\int \! {T_{st}^{n}\, dx_s}\right)}T_s^{n+1}T_t^{n+1}
\end{equation}
Notice that, unlike the original form of LBP, operations on graph neighborhoods $\prod_{u\in N(s)}$ do not differentiate between targets.

\section{Neural implementation of TRP updates}
\label{sec:NeuralTRP}

\subsection{Updating natural parameters}

TRP's operation only requires updating pseudomarginals, in place, using local information. These are appealing properties for a candidate brain algorithm. This representation is also nicely compatible with the structure of PPCs: different projections of the neural activity encode the natural parameters of an exponential family distribution. It is thus useful to express the pseudomarginals and the TRP inference algorithm using vectors of sufficient statistics $\vphi_c(\vx_c)$ and natural parameters $\vtheta_c^n$ for each clique: $T^n_c(\vx_c)=\exp{\left(\vtheta_c^n\cdot\vphi_c(\vx_c)\right)}$. For a model with at most pairwise interactions, the TRP updates (\ref{BPTRP}) can be expressed in terms of these natural parameters as
\begin{equation}
\label{ParamUpdates}
\vtheta_s^{n+1} = (1-d_s)\vtheta_s^n +\!\! \sum_{u\in N(s)} \!\!\bg_V(\vtheta_{su}^n)
\hspace{3em}
\vtheta_{st}^{n+1} = \vtheta_{st}^n  + Q_s \vtheta_s^{n+1} + Q_t \vtheta_t^{n+1} + \bg_E(\vtheta_{st}^n)
\end{equation}
where $d_s$ is the number of neighbors of $s$, and $Q_s$, $\bg_{V}$ and $\bg_{E}$ are matrices and nonlinear functions (for vertices $V$ and edges $E$) that are determined by the particular graphical model (see below). Since the natural parameters reflect log-probabilities, the product rule for probabilities becomes a linear sum in $\vtheta$, while the sum rule for probabilities must be implemented by nonlinear operations $\bg$ on $\vtheta$.

In the concrete case of a Gaussian graphical model, the joint distribution is given by $p(\vx) \propto \exp{ (-\frac{1}{2}\vx^\top \! \bA \vx  + \bb^\top\! \vx )}$, where $\bA$ and $\bb$ are the natural parameters, and the linear and quadratic functions $\vx$ and $\vx\vx^\top$ are the sufficient statistics. When we reparameterize this distribution by pseudomarginals, we again have linear and quadratic sufficient statistics: two for each node, $\vphi_s=(-\frac{1}{2}x_s^2,\ x_s)^\top$, and five for each edge, $\vphi_{st}=(-\frac{1}{2}x_s^2,\ x_sx_t,\ -\frac{1}{2}x_t^2,\ x_s,\ x_t)^\top$. Each of these vectors of sufficient statistics has its own vector of natural parameters, $\vtheta_s$ and $\vtheta_{st}$.

To approximate the marginal probabilities, the TRP algorithm initializes the pseudomarginals to $\vtheta_s^0 = \lpr A_{ss},~b_s\rpr^\top $ and 
$\vtheta_{st}^0 = \lpr A_{ss},~A_{st},~A_{tt},~b_s,~b_t\rpr^\top $. To update $\vtheta$, we must extract the matrices $Q$ and nonlinear functions $\bg$ that recover the univariate marginal distribution of a bivariate gaussian $T_{st}$. For $T_{st}(x_s,x_t) \propto \exp \lpr -\frac{1}{2}\theta_1 x_s^2 - \theta_2 x_s x_t -\frac{1}{2}\theta_3 x_t^2 + \theta_4 x_s + \theta_5 x_t \rpr$, this marginal is 
\begin{equation}
\label{BiGauss}
T_s(x_s) =\int dx_t\,T_{st}(x_s,x_t)\propto \exp \lpr -\frac{\theta_1 \theta_3 - \theta_2^2}{\theta_3}\frac{x_s^2}{2} + \frac{\theta_4 \theta_3 - \theta_2\theta_5}{\theta_3}x_s\rpr
\end{equation}
Using this, we can determine the form of the weight matrices and the nonlinear functions in the TRP updates (\ref{ParamUpdates}).
\begin{align}
\label{LinearComps}
Q_{s}&= \begin{pmatrix}
		1 & 0 & 0 & 0 & 0 \\
		0 & 0 & 0 & 1 & 0 \\
	\end{pmatrix}^\top
\hspace{2em}
Q_{t}= \begin{pmatrix}
	0 & 0 & 1 & 0 & 0 \\
	0 & 0 & 0 & 0 & 1 \\
\end{pmatrix}^\top\\
\bg_{V}(\vtheta_{su}^n) &= \lpr \frac{\vtheta_{1;su}^n \vtheta_{3;su}^n - \lpr\vtheta_{2;su}^n\rpr^2}{\vtheta_{3;su}^n},~\frac{\vtheta_{4;su}^n \vtheta_{3;su}^n - \vtheta_{2;su}^n \vtheta_{5;su}^n}{\vtheta_{3;su}^n} \rpr^\top  \\ \nonumber
	\bg_{E}(\vtheta_{st}^n) &= \lpr \frac{\lpr \vtheta_{2;st}^n \rpr^2}{\vtheta_{3;st}},~0,~\frac{\lpr \vtheta_{2;st}^n \rpr^2}{\vtheta_{1;st}},~\frac{\vtheta_{2;st}\vtheta_{5;st}}{\vtheta_{3;st}},~\frac{\vtheta_{2;st}\vtheta_{4;st}}{\vtheta_{1;st}}\rpr^\top 
\end{align}
where $\vtheta_{i;st}$ is the $i^{\textrm{th}}$ element of $\vtheta_{st}$. Notice that these nonlinear functions are all quadratic functions with a linear divisive normalization.

\subsection{Separation of Time Scales for TRP Updates}

An important feature of the TRP updates is that they circumvent the `message exclusion' problem of LBP. The TRP update for the singleton terms, (\ref{BPTRP}) and (\ref{ParamUpdates}), includes contributions from \textit{all the neighbors} of a given node. There is no free lunch, however, and the price is that the updates at time $n+1$ depend on previous pseudomarginals at two different times, $n$ and $n+1$. The latter update is therefore instantaneous information transmission, which is not biologically feasible.

To overcome this limitation, we observe that the brain can use fast and slow timescales $\tau_{\rm fast} \ll \tau_{\rm slow}$ instead of instant and delayed signals. We convert the update equations to continuous time, and introduce auxiliary variables $\tilde{\vtheta}$ which are lowpass-filtered versions of $\vtheta$ on a slow timescale: $\tau_{\rm slow} \dot{\tilde{\vtheta}} = -\tilde{\vtheta} + \vtheta$. The nonlinear dynamics of (\ref{ParamUpdates}) are then updated on a faster timescale $\tau_{\rm fast}$ according to
\begin{equation}
\label{ContinuousParamUpdates}
\tau_{\rm fast}\dot{\vtheta}_s = -d_s\tilde{\vtheta}_s +\!\! \sum_{u\in N(s)} \!\!\bg_V(\tilde{\vtheta}_{su})
\hspace{3em}
\tau_{\rm fast}\dot{\vtheta}_{st} = Q_s \vtheta_s + Q_t \vtheta_t + \bg_E(\tilde{\vtheta}_{st})
\end{equation}
where the nonlinear terms $\bg$ depend only on the slower, delayed activity $\tilde{\vtheta}$. By concatenating these two sets of parameters, $\vTheta = (\vtheta,\tilde{\vtheta})$, we obtain a coupled multidimensional dynamical system which represents the approximation to the TRP iterations:
\begin{equation}
\label{DSEq}
	\dot{\vTheta} = W\vTheta + \bG(\vTheta)
\end{equation}
Here the weight matrix $W$ and the nonlinear function $\bG$ inherit their structure from the discrete-time updates and the lowpass filtering at the fast and slow timescales.

\subsection{Network Architecture}
\label{NNet}

To complete our neural inference network, we now embed the nonlinear dynamics (\ref{DSEq}) into the population activity $\br$. Since different projections of the neural activity in a linear PPC encode natural parameters of the underlying distribution, we map neural activity onto $\vTheta$ by
\begin{equation}
\br = U \vTheta
\label{eq:projection}
\end{equation}
where $U$ is a rectangular $N_\br\times N_\vTheta$ embedding matrix that projects the natural parameters and their low-pass versions into the neural response space. These parameters can be decoded from the neural activity as $\vTheta = U^{+}\br$, where $U^{+}$ is the pseudoinverse of $U$.

\begin{figure}[h]
	\centering
	\includegraphics[width=0.9\textwidth]{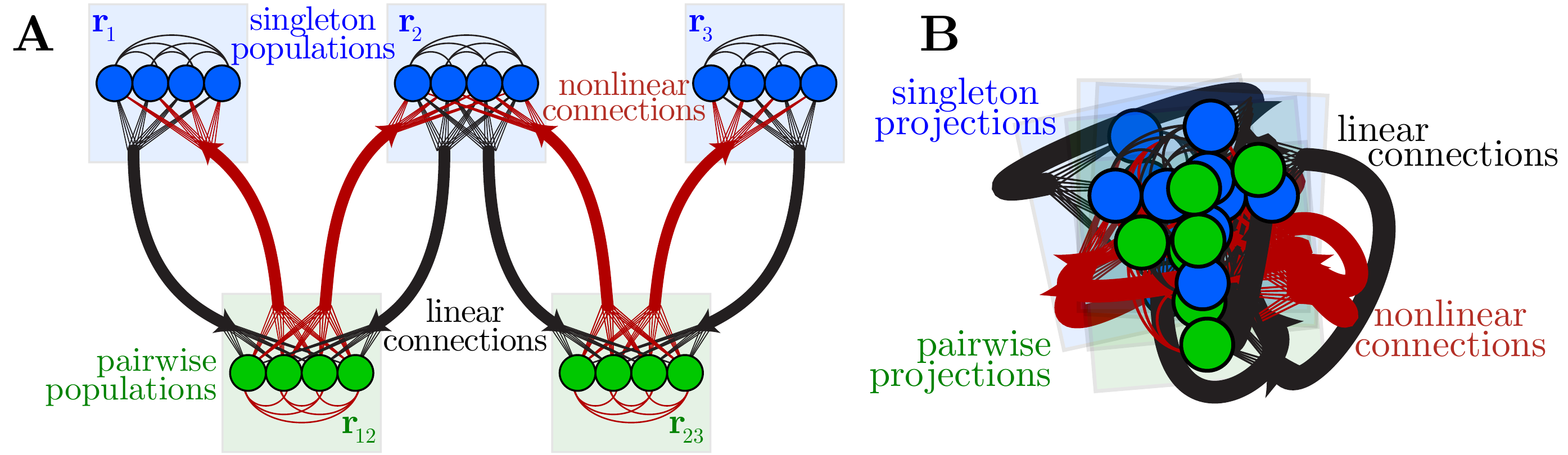}
	\caption{Distributed, nonlinear, recurrent network of neurons that performs probabilistic inference on a graphical model. ({\bf A}) This simple case uses distinct subpopulations of neurons to represent different factors in the example model in Figure \ref{fig:TRP2}A. ({\bf B}) A cartoon shows how the same distribution can be represented as distinct projections of the distributed neural activity, instead of as distinct populations. In both cases, since the neural activities encode log-probabilities, linear connections are responsible for integrating evidence while nonlinear connections perform marginalization.}
	\label{NN}
\end{figure}

Applying this basis transformation to (\ref{DSEq}), we have $\dot{\br} = U \dot{\vTheta}=U(W\vTheta+\bG(\vTheta))=UW U^{+} \br + U\bG(\bU^+\br)$. We then obtain the general form of the updates for the neural activity
\begin{equation}
\label{eq:NeuralTRP}
\dot{\br} = W_{\!L}\br + \bG_{\!NL}(\br)
\end{equation}
where $W_L\br=UWU^+\br$ and $\bG_{NL}(\br)=U\bG(U^+\br)$ correspond to the linear and nonlinear computational components that integrate and marginalize evidence, respectively. The nonlinear function on $\br$ inherits the structure needed for the natural parameters, such as a quadratic polynomial with a divisive normalization used in low-dimensional Gaussian marginalization problems \cite{beck2011marginalization}, but now expanded to high-dimensional graphical models. Figure \ref{NN} depicts the network architecture for the simple graphical model from Figure \ref{fig:TRP2}A, both when there are distinct neural subpopulations for each factor (Figure \ref{NN}A), and when the variables are fully multiplexed across the entire neural population (Figure \ref{NN}B). These simple, biologically-plausible neural dynamics (\ref{eq:NeuralTRP}) represent a powerful, nonlinear, fully-recurrent network of PPCs which implements the TRP update equations on an underlying graphical model.

 \section{Experiments}
 \label{Experiments} 
 
\begin{figure}[h]
 	\centering
 	\includegraphics[width=0.8\textwidth]{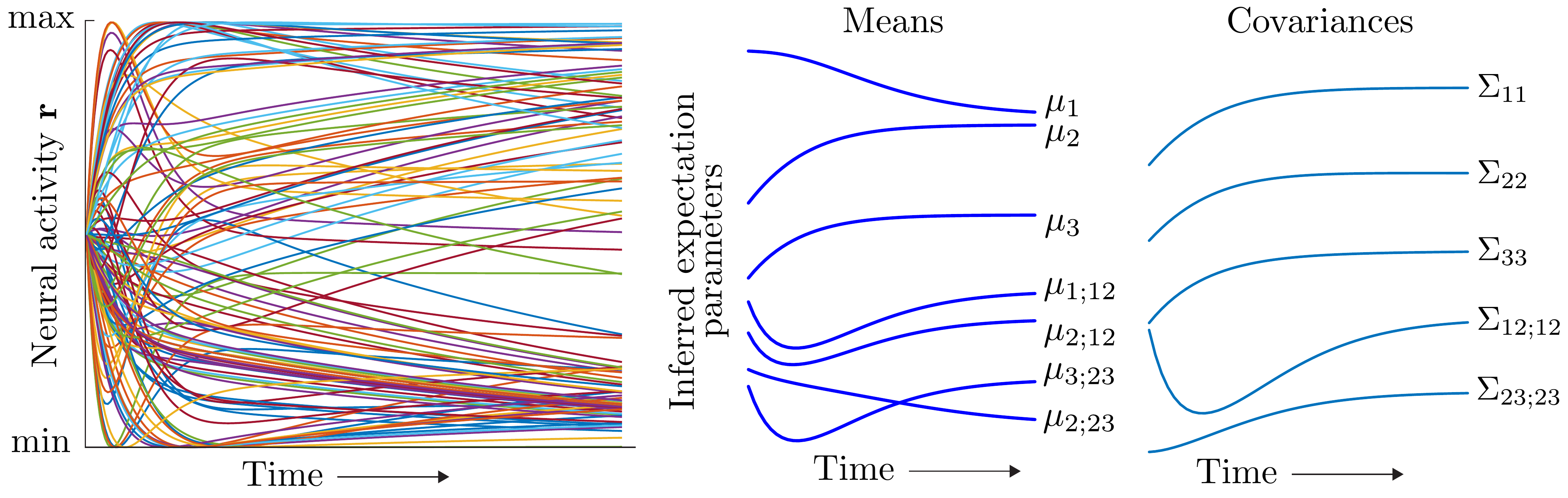}
 	\caption{Dynamics of neural population activity (left) and the expectation parameters of the posterior distribution that the population encodes (right) for one trial of the tree model in Figure \ref{fig:TRP2}A.}
 	\label{fig:dynamics}
 \end{figure}

We evaluate the performance of our neural network on a set of small Gaussian graphical models with up to 400 interacting variables. The networks time constants were set to have a ratio of $\tau_{\rm slow}/\tau_{\rm fast}=20$. Figure \ref{fig:dynamics} shows the neural population dynamics as the network performs inference, along with the temporal evolution of the corresponding node and pairwise means and covariances. The neural activity exhibits a complicated timecourse, and reflects a combination of many natural parameters changing simultaneously during inference. This type of behavior is seen in neural activity recorded from behaving animals \cite{hayden2010neurons,rigotti2013importance,raposo2014category}.

\begin{figure}[h]
 	\centering
 	\includegraphics[width=.9\textwidth]{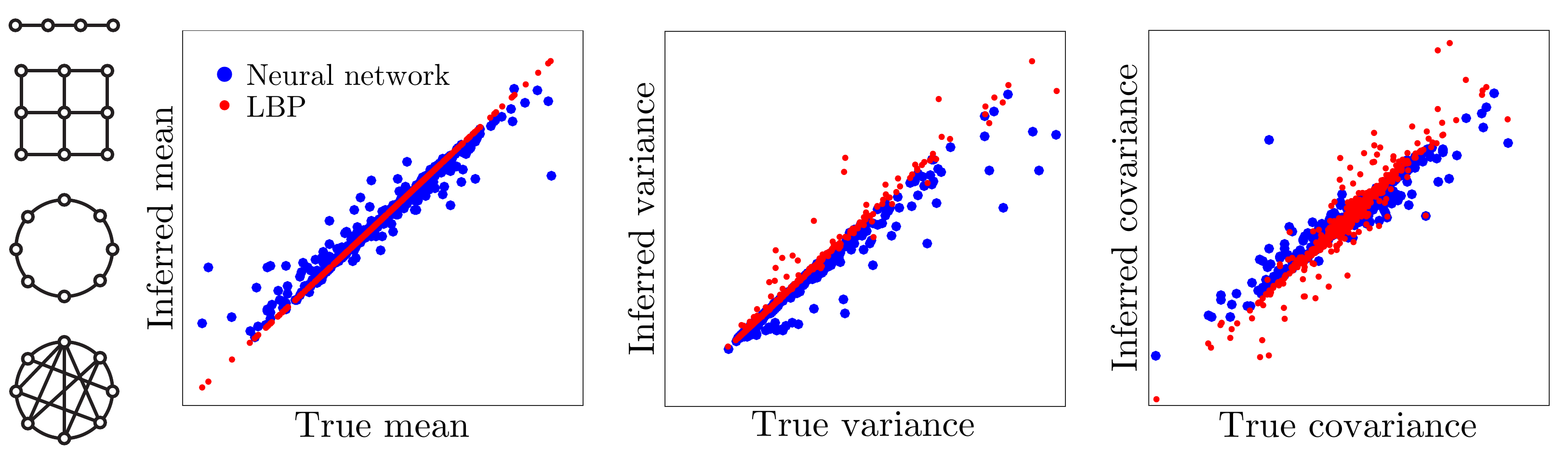}
 	\caption{Inference performance of our neural network (blue) and standard loopy belief propagation (red) for a variety of graph topologies, including square grids up to $20\times 20$ and densely connected graphs with up to 25 variables. The expectation parameters (means, covariances) of the pseudomarginals closely match the corresponding parameters for the true marginals.}
 	\label{fig:performance}
 \end{figure}

\begin{figure}[h]
  	\centering
  	\includegraphics[width=1\textwidth]{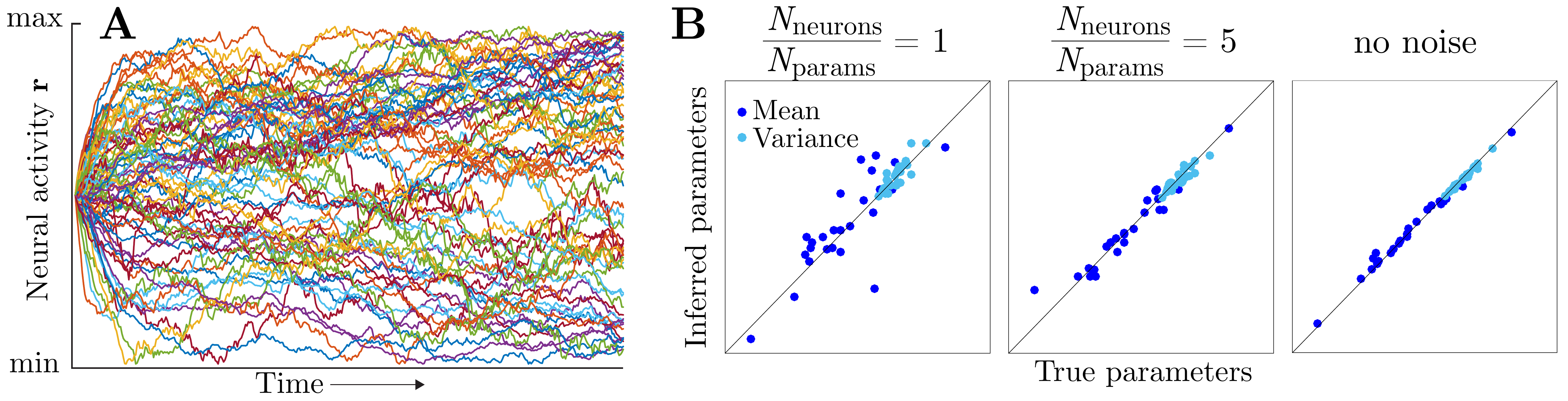}
  	\caption{Network performance is robust to noise, and improves with more neurons. ({\bf A}) Neural activity performing inference on a $5\times 5$ square grid, in the presence of independent spatiotemporal Gaussian noise of standard deviation 0.1 times the standard deviation of each signal. ({\bf B}) Expectation parameters (means, variances) of the node pseudomarginals closely match the corresponding parameters for the true marginals, despite the noise. Results are shown for one or five neurons per parameter in the graphical model, and for no noise (i.e. infinitely many neurons).}
  	\label{fig:noiseperformance}
  \end{figure}

Figure \ref{fig:performance} shows that our recurrent neural network accurately infers the marginal probabilities, and reaches almost the same conclusions as loopy belief propagation. The data points are obtained from multiple simulations with different graph topologies, including graphs with many loops. Figure \ref{fig:noiseperformance} verifies that the network is robust to noise even when there are few neurons per inferred parameter; adding more neurons improves performance since the noise can be averaged away.
 
 \section{Conclusion}
 \label{Conclusion}
 
We have shown how a biologically-plausible nonlinear recurrent network of neurons can represent a multivariate probability distribution using population codes, and can perform inference by reparameterizing the joint distribution to obtain approximate marginal probabilities.

Our network model has desirable properties beyond those lauded features of belief propagation. First, it allows for a thoroughly distributed population code, with many neurons encoding each variable and many variables encoded by each neuron. This is consistent with neural recordings in which many task-relevant features are multiplexed across a neural population \cite{hayden2010neurons,rigotti2013importance,raposo2014category}.

Second, the network performs inference in place, without using a distinct neural representation for messages, and avoids the biological implausibility associated with sending different messages about every variable to different targets. This virtue comes from exchanging multiple messages for multiple timescales. It is noteworthy that allowing two timescales prevents overcounting of evidence on loops of length two (target to source to target). This suggests a novel role of memory in static inference problems: a longer memory could be used to discount past information sent at more distant times, thus avoiding the overcounting of evidence that arises from loops of length three and greater. It may therefore be possible to develop reparameterization algorithms with all the convenient properties of LBP but with improved performance on loopy graphs.
 
Previous results show that the quadratic nonlinearity with divisive normalization is convenient and biologically plausible interpretable, but this precise form is not necessary: other pointwise neuronal nonlinearities are capable of producing the same high-quality marginalizations in PPCs \cite{vasudeva2015marginalization}. In a distributed code, the precise nonlinear form at the neuronal scale is not important as long as the effect on the parameters is the same.

More generally, however, different nonlinear computations on the parameters implement different approximate inference algorithms. The distinct behaviors of such algorithms as mean-field inference, generalized belief propagation, and others arise from differences in their nonlinear transformations. Even Gibbs sampling can be described as a noisy nonlinear message-passing algorithm. Although LBP and its generalizations have strong appeal, we doubt the brain will use this algorithm exactly. The real nonlinear functions in the brain may implement even smarter algorithms.

To identify the brain's algorithm, it may be more revealing to measure how information is represented and transformed in a low-dimensional latent space embedded in the high-dimensional neural responses than to examine each neuronal nonlinearity in isolation. The present work is directed toward this challenge of understanding computation in this latent space. It provides a concrete example showing how distributed nonlinear computation can be distinct from localized neural computations. Learning this computation from data will be a key challenge for neuroscience. In future work we aim to recover the latent computations of our network from artificial neural recordings generated by the model. Successful model recovery would encourage us to apply these methods to large-scale neural recordings to uncover key properties of the brain's distributed nonlinear computations.

\newcommand{\RR}{RR }
\newcommand{\XP}{XP }

\subsubsection*{Author contributions} \XP conceived the study. \RR and \XP derived the equations. \RR implemented the computer simulations. \RR and \XP analyzed the results and wrote the paper.

{\bf Acknowledgments:} \XP and \RR were supported by a grant from the McNair Foundation and by the Intelligence Advanced Research Projects Activity (IARPA) via Department of Interior/Interior Business Center (DoI/IBC) contract number D16PC00003.\footnote{The U.S. Government is authorized to reproduce and distribute reprints for Governmental purposes notwithstanding any copyright annotation thereon. Disclaimer: The views and conclusions contained herein are those of the authors and should not be interpreted as necessarily representing the official policies or endorsements, either expressed or implied, of IARPA, DoI/IBC, or the U.S. Government.}

\subsection*{References}
\small
\label{sec:references}
\bibliographystyle{plos.bst}
{\def\section*#1{}
\bibliography{Inference}
}

\end{document}